\documentstyle[psfig]{article}

\advance\textwidth by 100pt
\advance\oddsidemargin by -50pt
\advance\evensidemargin by -50pt

\begin{document}

\title{Constraints on the mSUGRA parameter\\ space from electroweak precision
data\footnote{Contribution of the Precision Electroweak Subgroup of the SUGRA 
Working Group for the Physics at Run~II -- Supersymmetry/Higgs Workshop,
November 19-21, 1998.}}
\author{G.C.~Cho$^1$, K.~Hagiwara$^2$, C.~Kao$^3$ and 
R.~Szalapski$^{4,}$\footnote{subgroup convener}}
\date{January, 1999}
\maketitle


\vspace*{-7cm}\hspace*{12cm}
\vbox{\baselineskip14pt
\hbox{\bf UR-1558}
\hbox{ER/40685/927}
\hbox{\bf KEK-TH-609}
}
\vspace*{5cm}

\begin{center}
$^1$Theory Group, KEK, Tsukuba, Ibaraki 305, Japan\\
e-mail: cho@theory.kek.jp\\[0.2cm]
$^2$Theory Group, KEK, Tsukuba, Ibaraki 305, Japan\\
e-mail: kaoru.hagiwara@kek.jp\\[0.2cm]
$^3$Department of Physics, University of Wisconsin -- Madison\\ 
 1150 University Avenue, Madison, WI 53706, USA\\
 e-mail: kao@pheno.physics.wisc.edu\\[0.2cm]
$^4$Department of Physics and Astronomy, University of Rochester\\
River Campus, B\&L Bldg., Rochester, NY 14627-0171, USA\\
E-mail: robs@pas.rochester.edu
\end{center}

\begin{abstract}
We place constraints on the parameter space of the minimal supergravity (SUGRA)
inspired supersymmetric (SUSY) extension of the standard model (SM), {\em i.e.}
the mSUGRA model, by studying the loop-level contributions of SUSY particles to
electroweak precision observables.  In general the Higgs bosons and the 
superpartner particles of SUSY models contribute to electroweak observables 
through universal propagator corrections as well as process-specific vertex and
box diagrams.  However, due to the bound on the mass of the lightest chargino, 
$m_{{\tilde{\chi}}_1^\pm} > 91$~GeV, we find that the process-dependent 
contributions to four-fermion amplitudes are negligibly small.  Hence, the full
analysis may be reduced to an analysis of the propagator corrections, and in 
some regions of parameter space the constraints from the $b\rightarrow s\gamma$
process are quite important.  The propagator corrections are dominated by the 
contributions of the scalar fermions, and we summarize the results in the 
Peskin-Takeuchi $S$--$T$ plane and the contributions to the $W$-boson mass, 
$m_W$.  We then present the results in the mSUGRA $m_0$--$m_{1/2}$ plane and 
find that our analysis of the propagator corrections provides constraints in 
the small-$m_0$--small-$m_{1/2}$ region, precisely the region of interest for 
collider phenomenology.  In some regions of parameter space, especially for 
$\mu < 0$ and large $\tan\beta$, the constrained region is enlarged 
considerably by including the process $b\rightarrow s\gamma$.
\end{abstract}


This is the report of the Electroweak Precision Working Subgroup of the 
SUGRA Working Group for the Physics at Run~II -- Supersymmetry/Higgs 
Workshop.  As such, we forgo the usual introduction and defer to the larger
working-group report.\cite{run2-susyhiggs}  Our task is to place constraints 
on the parameter space of the minimal supergravity (SUGRA) inspired 
supersymmetric (SUSY) extension of the standard model (SM), {\em i.e.} the 
mSUGRA model, by studying the loop-level contributions of the supersymmetric 
particles to electroweak precision observables.  The work presented here is 
part of a  larger collaborative effort, and results will be presented more 
completely elsewhere.\cite{chks99}  


The loop-level contributions of supersymmetric (SUSY) particles to electroweak
observables have been extensively discussed in the 
literature.\cite{susy_loop1,susy_loop2,susy_loop3,susy_loop4}  In particular,
processes with four external light fermions have been studied including 
observables  which are sensitive to the $Zbb$ coupling.  The branching 
fraction ${\rm Br}(B\rightarrow X_s\gamma)$ is sensitive to 
SUSY effects in some regions of parameter space.\cite{bsg-theory,more-bsg}
The relationship between $m_W$ and $m_Z$ will provide stronger constraints
as the measurement of $m_W$ improves.

The complete one-loop corrections to four-fermion amplitudes include the 
universal propagator corrections as well as the process-dependent vertex 
and box corrections.  However, when the extra Higgs bosons and the superpartner
particles become sufficiently massive, it is necessary to retain only the 
leading propagator corrections\cite{susy_loop_heavy}, and these contributions 
may be summarized in terms of the $S$, $T$ and $U$ parameters of Peskin and 
Takeuchi\cite{stu} or some other triplet of parameters.\cite{others}  The 
recent bounds\cite{susy_bound} on the mass of the lightest chargino,  
$m_{\tilde{\chi}^\pm_1} > 91$~GeV, and on the mass of the lighter scalar-top 
quark, $m_{\tilde{t}_1} > 80$~GeV, imply a sufficiently massive spectrum such 
that the process-dependent vertex and box contributions may be safely 
neglected.  In the context of the mSUGRA model, the chargino mass bound alone 
is sufficient to reach this conclusion.

In our analysis we adopt, in the notation of Hagiwara {\em et al.}\cite{hhkm}, 
a form factor, $g_L^b$, to describe corrections to the $Zbb$ vertex as well 
as the $S$ and $T$ parameters which include corrections to the gauge-boson 
propagators.  We find that it is more convenient to drop the $U$ 
parameter in favor of the directly measured $W$-boson mass.  We first obtain 
constraints from the electroweak data on the four parameters $\Delta g_L^b$, 
$\Delta S$, $\Delta T$ and $\Delta m_W$, which measure deviations from their 
corresponding SM reference values calculated at $m_t = 175$~GeV and 
$m_H = 100$~GeV.  We then calculate the contributions to these parameters 
and to the $B\rightarrow X_s\gamma$ decay width from the superpartner and 
Higgs particles to obtain constraints on the mSUGRA parameters.

The electroweak data through 1998 including the LEP 
and SLC experiments as well as low-energy neutral-current experiments may 
be summarized as 
\begin{equation}
\left.
\begin{array}{ccc}
\Delta S - 24.2 \Delta g_L^b & = & -0.114 \pm 0.14 \\
\Delta T - 42.9 \Delta g_L^b & = & -0.215 \pm 0.14 
\end{array}
\right\}\makebox[0.4cm]{}
\rho_{\rm corr} = 0.77 \;,
\end{equation}
where $\rho_{\rm corr}$ denotes the correlation between the two one-sigma 
errors.  Because the correlation is strong we present our results in the 
$\Delta S^\prime$--$\Delta T^\prime$ plane where $\Delta S^\prime = \Delta 
S - 24.2 \Delta g_L^b$ and $\Delta T^\prime = \Delta T - 42.9 \Delta 
g_L^b$.  
Note that $m_W$ is not correlated with $\Delta S^\prime$ and $\Delta T^\prime$,
and hence it may be treated separately.  Averaging the LEP2 and Tevatron 
measurements of the $W$-boson mass,  $m_W = 80.375 \pm 0.064$~GeV.
The deviation of the data from the SM reference value for the $W$-boson mass is
\begin{equation}
\Delta m_W = -0.027 \pm 0.064 \rm GeV \;. 
\end{equation}
For the measurement of the branching fraction for the process $b\rightarrow
s\gamma$ we use
\begin{equation}
{\rm Br}(B\rightarrow X_s\gamma) = 
3.11 \pm 0.80 \pm 0.72 \times 10^{-4}\;,
\end{equation}
from the ALEPH\cite{bsg-aleph} collaboration.  Results from the more recent 
CLEO measurement\cite{bsg-cleo} will be reported elsewhere\cite{chks99}.


The SUSY contributions to $\Delta S^\prime$, $\Delta T^\prime$ and 
$\Delta m_W$ are dominated by the contributions of the sfermions.
Hence, we begin with a discussion of the sfermion contributions.
\begin{figure}[htb]
\begin{center}
\leavevmode\psfig{figure=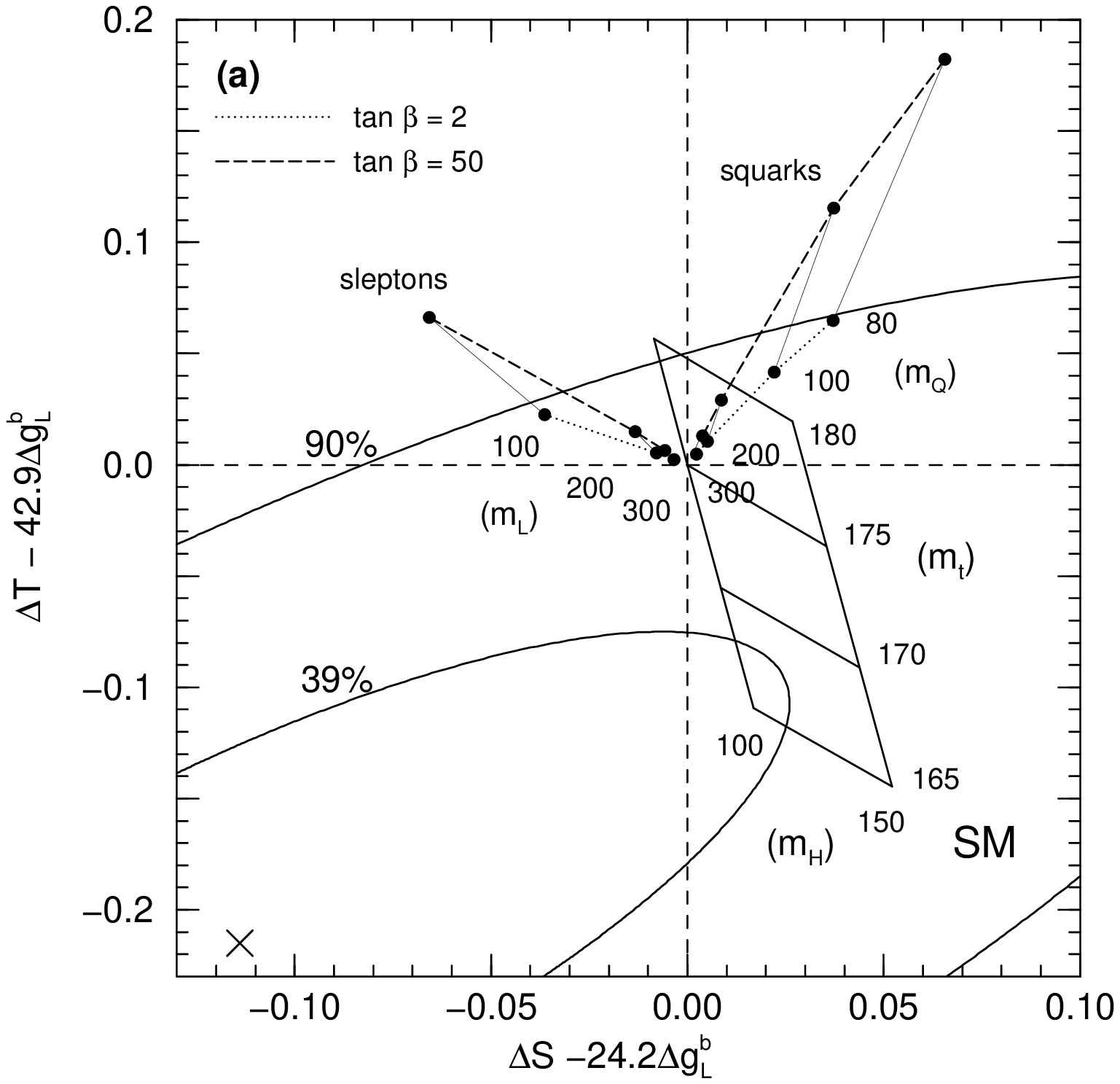,width=8.5cm,silent=0}
\leavevmode\hspace*{0.5cm}
\leavevmode\psfig{figure=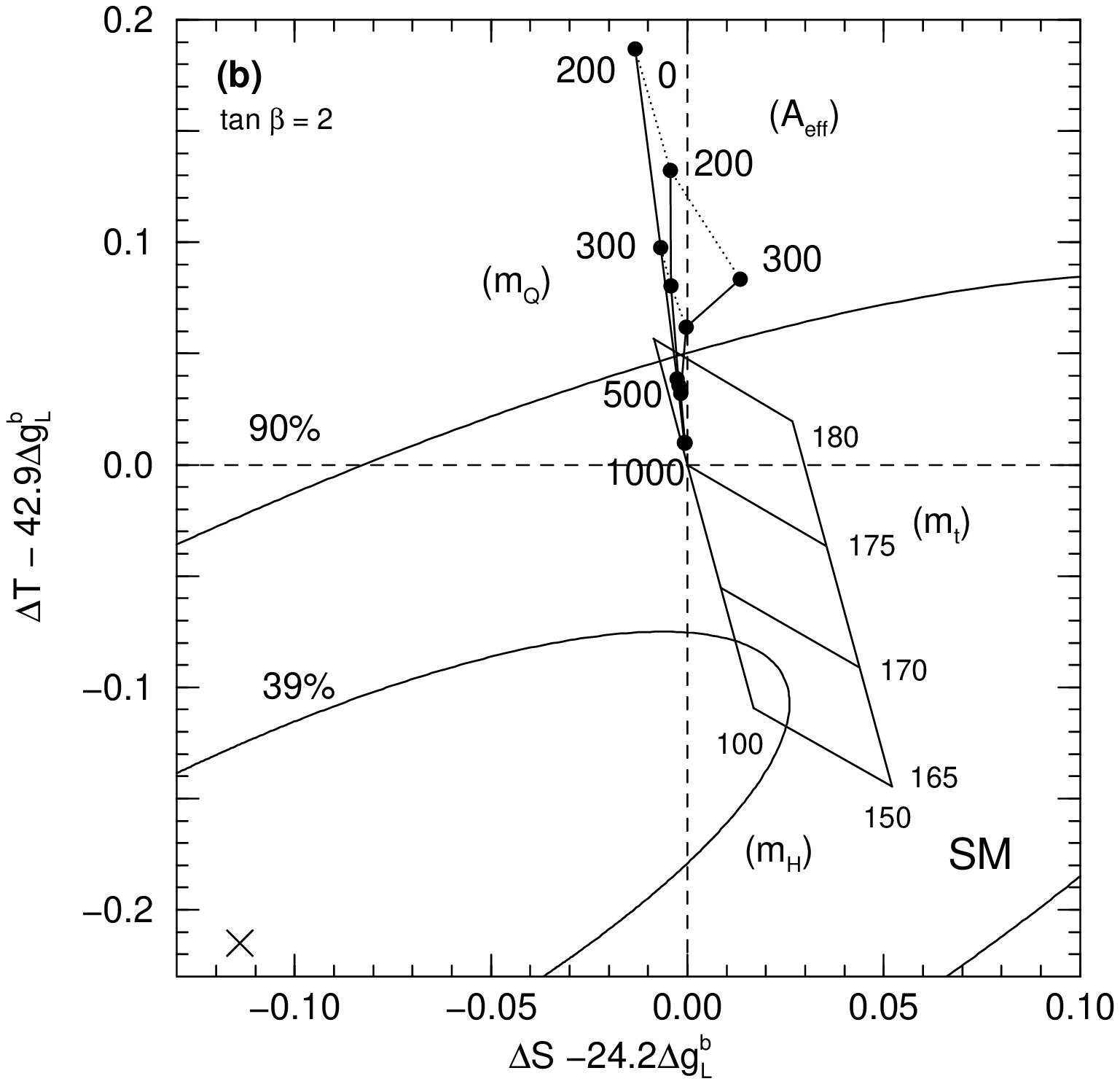,width=8.5cm,silent=0}
\end{center}
\caption{(a) shows the sfermion contributions for the first two families,
and (b) shows the stop-sbottom contributions.  Details are given in the text.}
\label{fig_s-t-plane}
\end{figure}
In Figure~\ref{fig_s-t-plane}(a) and (b) the `$\times$' marks the location of 
the best fit to the experimental data in the 
$\Delta S^\prime - \Delta T^\prime$
plane, and the ellipses show the 39\% (one-sigma) and 90\% confidence-level 
(CL) contours as
indicated.  A grid has been included which shows the SM predictions for 
$\Delta S^\prime$ and $\Delta T^\prime$ as a function of $m_t$ and $m_H$.
We choose the point where $m_t = 175$~GeV and $m_H = 100$~GeV as our reference
point, {\em i.e.} $\Delta S^\prime = \Delta T^\prime = 0$, and the dashed-line
axes are drawn through this point. The same point serves as the SUSY prediction
in the limit of very large masses for the non-SM particles and when the 
lightest SUSY Higgs particle behaves like the SM Higgs boson.

Figure~\ref{fig_s-t-plane}(a) includes the contributions of the sfermions of 
the first two generations with the squark and slepton contributions 
shown separately.  The contribution of a sfermion loop to the $S$ parameter is
proportional to the hypercharge of the sfermion.  Since $Y = \frac{1}{6}$ for 
the squarks and $Y = -\frac{1}{2}$ for the sleptons, we see that the 
squarks increase $\Delta S^\prime$ while the sleptons decrease 
$\Delta S^\prime$.  Dotted contours are used to show the case where 
$\tan\beta = 2$ while dashed contours are used to show the $\tan\beta = 50$ 
case.  For the slepton contributions we show the cases where the explicit 
soft-SUSY-breaking slepton-doublet mass parameter has the nonzero values 
$m_L = 100, 200$ and 300~GeV.  Contours of equal $m_L$ but varying 
$\tan\beta$ are drawn using thin solid lines.  Similarly we consider the 
squark contributions where the explicit soft-SUSY-breaking squark-doublet mass 
parameter has the values $m_Q = 80, 100, 200$ and 300~GeV; contours of 
constant $m_Q$ but varying $\tan\beta$ are indicated by the thin solid lines.
While the contributions to $\Delta S^\prime$ tend to cancel between the 
squark and sfermion sectors, the contributions to $\Delta T^\prime$ always 
add constructively, and for light sfermions lead to an unacceptably large
deviation from the SM prediction and the experimental measurement of  
$\Delta T^\prime$.

The large mass of the top quark leads to large left-right mixing of the 
top squarks, and to a lesser degree the mass of the bottom quark leads to 
left-right mixing of the bottom squarks.  For this reason the third-family
sfermions require a separate discussion, and we summarize the stop--sbottom
contributions in Figure~\ref{fig_s-t-plane}(b).  In the mass matrix for the 
stop squarks it is the off-diagonal element $-m_t A_{\rm eff}^t$ where 
$A_{\rm eff}^t = A_t + \mu\cot\beta$ that determines the level of left-right 
mixing, while in the sbottom-squark mass matrix the off-diagonal element 
$-m_b A_{\rm eff}^b$ where $A_{\rm eff}^b = A_b + \mu\tan\beta$ determines
the degree of mixing.  We plot our results for $A_{\rm eff}^t = A_{\rm eff}^b
= A_{\rm eff}$ showing contours of constant $A_{\rm eff}$ by the dashed lines
and lines of constant $m_Q$ by the dotted lines.  In 
Figure~\ref{fig_s-t-plane}(a) we saw that, with a value as small as 
$m_Q = 80$~GeV, the contributions of the squarks of the first two generations 
to $\Delta T^\prime$ are still fairly small, while for the third family a 
value of $m_Q = 300$~GeV already produces an unacceptable result for 
reasonable  values of $A_{\rm eff}$.  It may be tempting to abandon
universality of the soft-SUSY-breaking parameters and consider cases with a
relatively small value of $m_Q$ for the first two families and a much larger 
value to decouple the third family.  While this is possible in principle, 
caution is required to avoid large flavor-changing neutral currents.  In the 
context of the mSUGRA model we will, of course, use the soft-SUSY-breaking 
parameters which are obtained from the common mass parameters at the GUT scale.
We also note that large values of $A_{\rm eff}$ tend to produce smaller 
$\Delta T^\prime$ but larger $\Delta S^\prime$.  We have shown only the case 
$\tan\beta = 2$ since we find similar results for large $\tan\beta$.


\begin{figure}[htbp]
\begin{center}
\leavevmode\psfig{figure=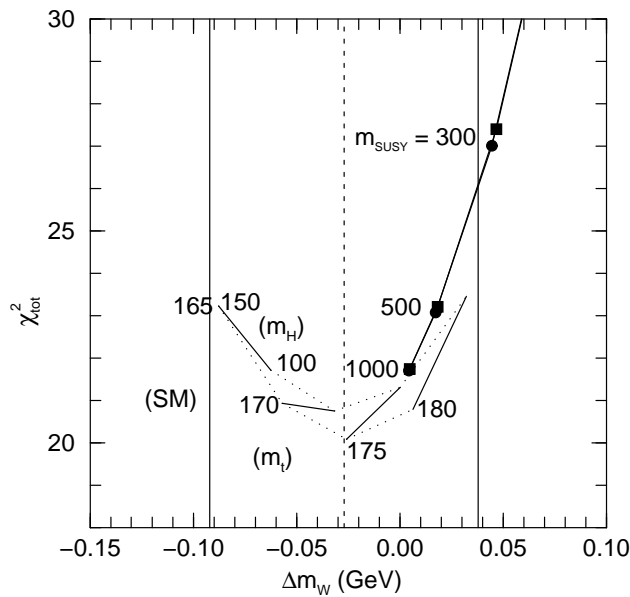,width=8.5cm,silent=0}
\end{center}
\caption{The sfermion contributions in the $\chi^2_{\rm tot} - \Delta m_W$
plane where $\chi^2_{\rm tot}$ refers to the total $\chi^2$ coming from the 
simultaneous fitting of $\Delta S^\prime$, $\Delta T^\prime$ and $\Delta m_W$.}
\label{fig_chi2-mw-plane}
\end{figure}
Figure~\ref{fig_chi2-mw-plane} shows the sfermion contributions to the 
$W$-boson mass.  We include a grid that shows the SM prediction for 
$\Delta m_W$ as a function of $m_H$ and $m_t$.  Along the upper dotted contour 
$m_H = 100$~GeV, while the lower dotted contour corresponds to $m_H = 150$~GeV.
Points of equal $m_t$ are connected by the solid line segments.  The vertical 
dashed line represents the world average for the central value of the $m_W$ 
measurement with the one-sigma errors represented by the vertical solid lines. 
For simplicity we set the explicit soft-SUSY-breaking squark-doublet, 
squark-singlet, slepton-doublet and slepton-singlet mass parameters to a 
common value, $m_{\rm SUSY}$.  We then plot the total chi-squared from the 
simultaneous fitting of $\Delta S^\prime$, $\Delta T^\prime$ and $\Delta m_W$, 
{\em i.e.} $\chi^2_{\rm tot}$, {\em versus} $\Delta m_W$ for $\tan\beta = 2$ 
(represented by the squares) and $\tan\beta = 50$ (represented by the circles).
For $m_{\rm SUSY} = 1000$~GeV the $\tan\beta = 2$ and $\tan\beta = 50$ points 
are nearly indistinguishable.  We note that the contributions of the SUSY 
particles always increase $m_W$.  However, a value of $m_{\rm SUSY} = 300$~GeV 
leads to only a one-sigma discrepancy with the data.  Hence, at the current 
time, the measurement of the $W$-boson mass provides only a minor constraint. 


Although the Higgs bosons, the charginos and the neutralinos also contribute 
to $\Delta S^\prime$, $\Delta T^\prime$ and $\Delta m_W$, in the mSUGRA model
the contributions are small compared to the sfermion contributions.  Hence, 
even though we include these contributions in the numerical analysis, we 
do not show the Higgs-boson, chargino and neutralino figures that correspond 
to Figure~\ref{fig_s-t-plane} and Figure~\ref{fig_chi2-mw-plane}.


\begin{figure}[htb]
\begin{center}
\leavevmode\psfig{figure=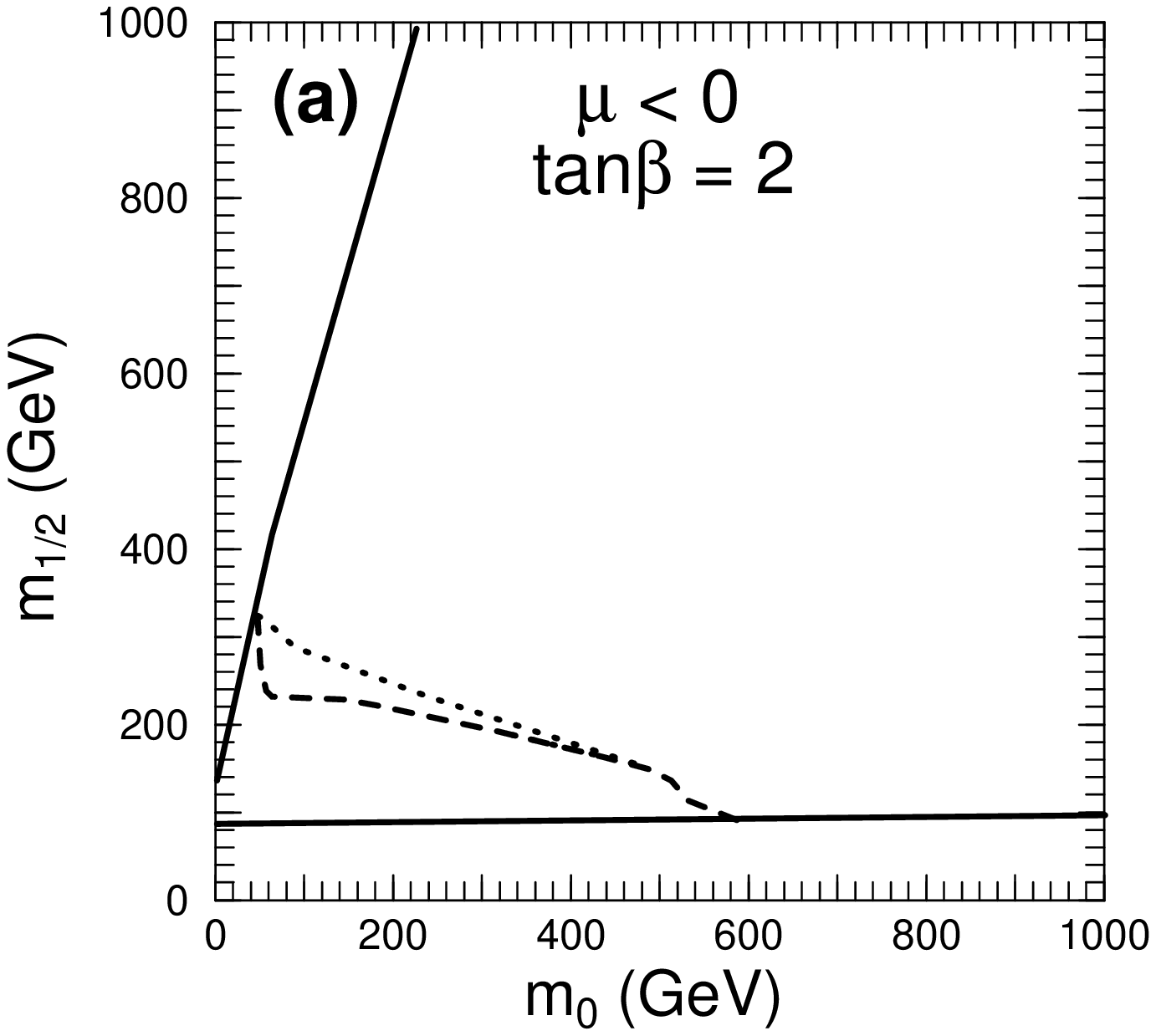,width=5.5cm,silent=0}
\leavevmode\hspace*{0.5cm}
\leavevmode\psfig{figure=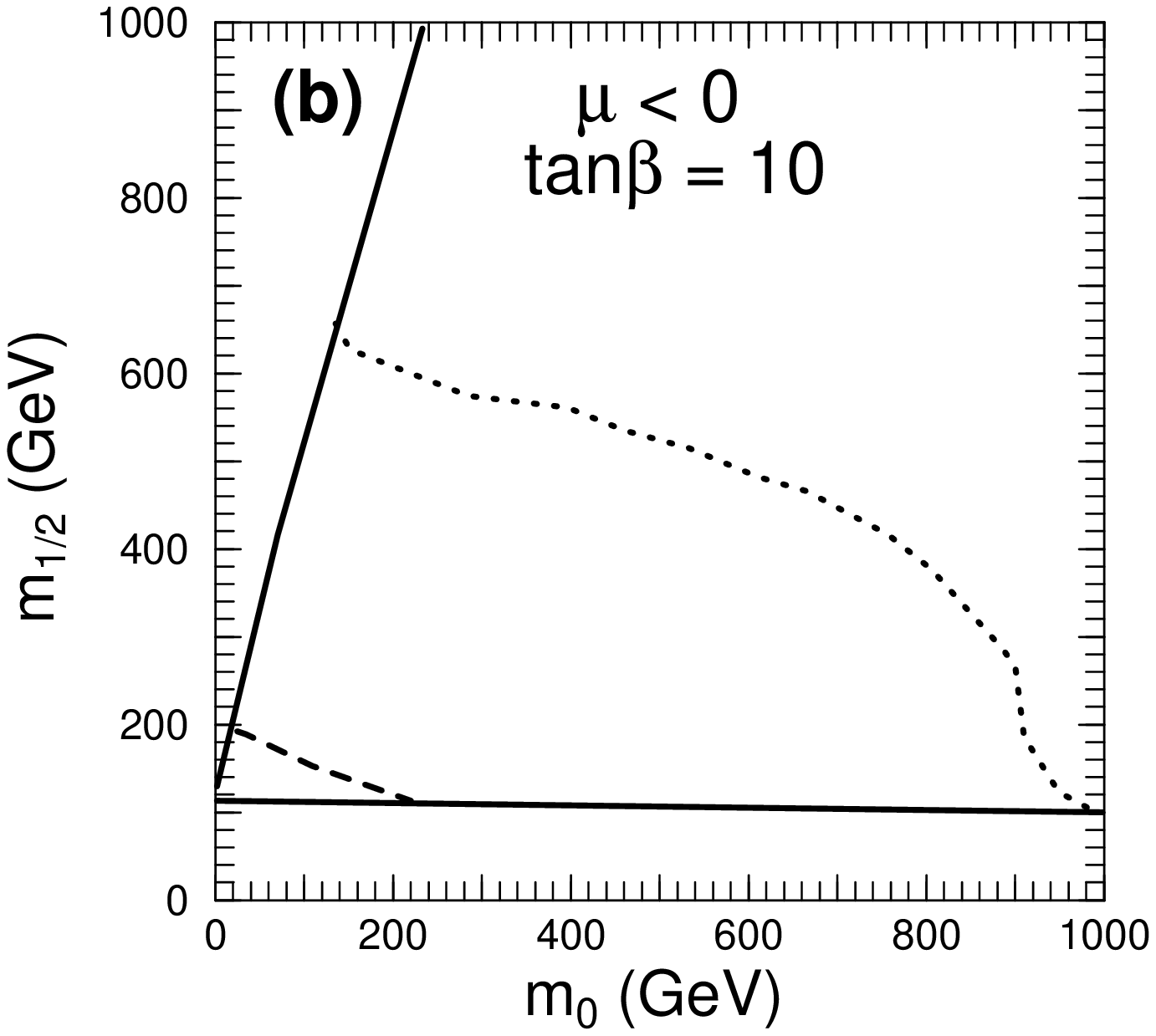,width=5.5cm,silent=0}
\leavevmode\hspace*{0.5cm}
\leavevmode\psfig{figure=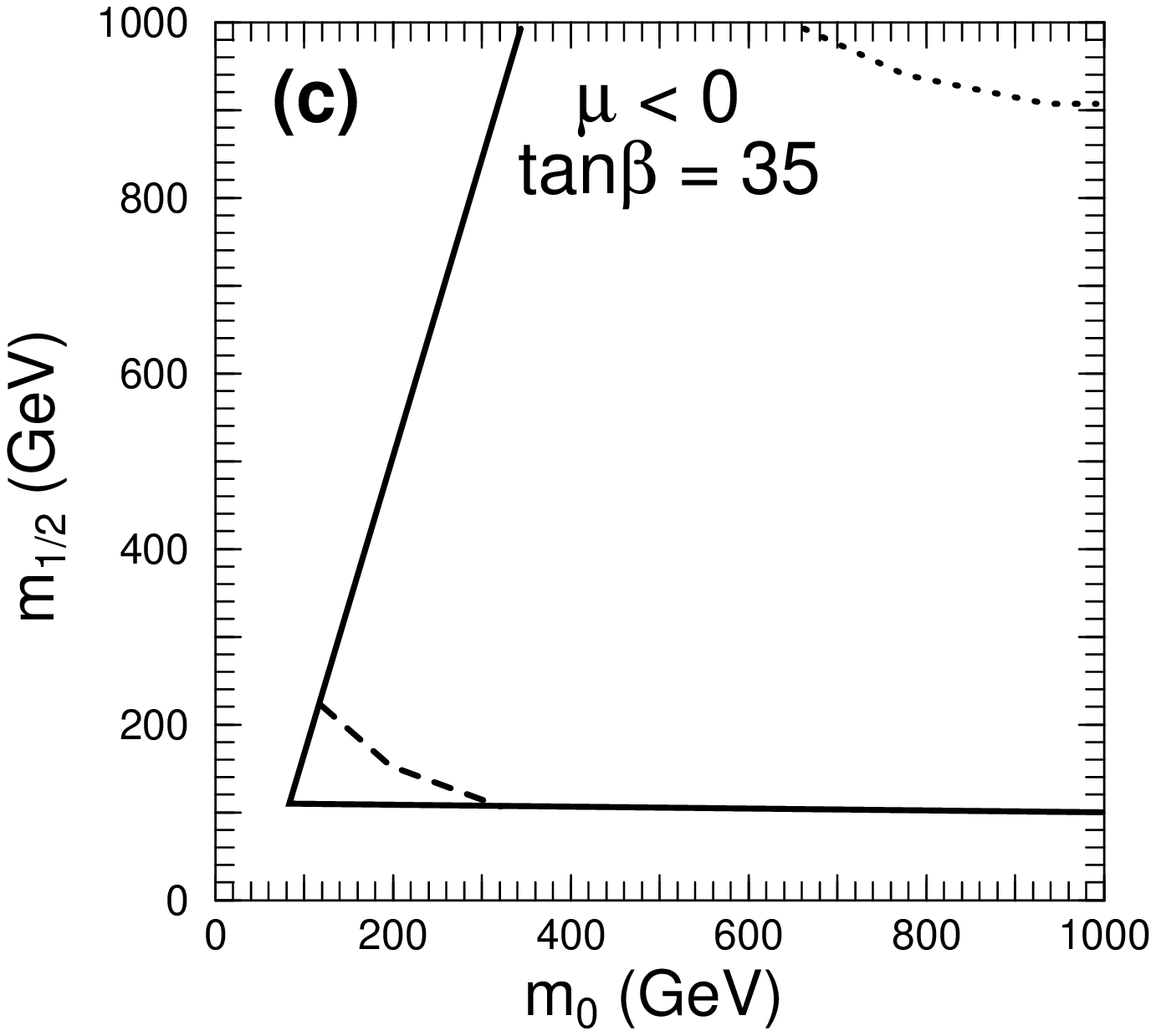,width=5.5cm,silent=0}
\end{center}
\vspace*{0.5cm}
\begin{center}
\leavevmode\psfig{figure=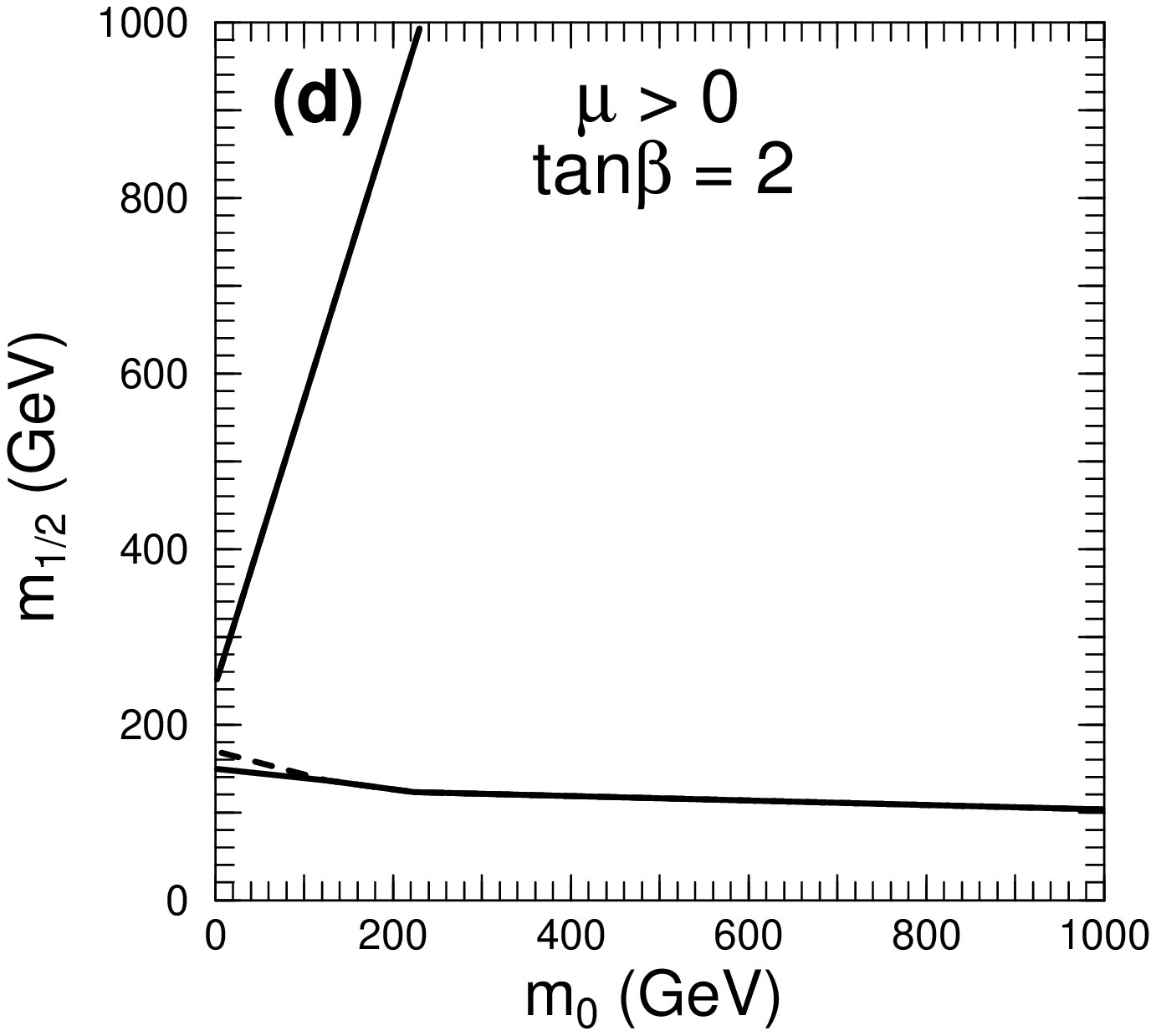,width=5.5cm,silent=0}
\leavevmode\hspace*{0.5cm}
\leavevmode\psfig{figure=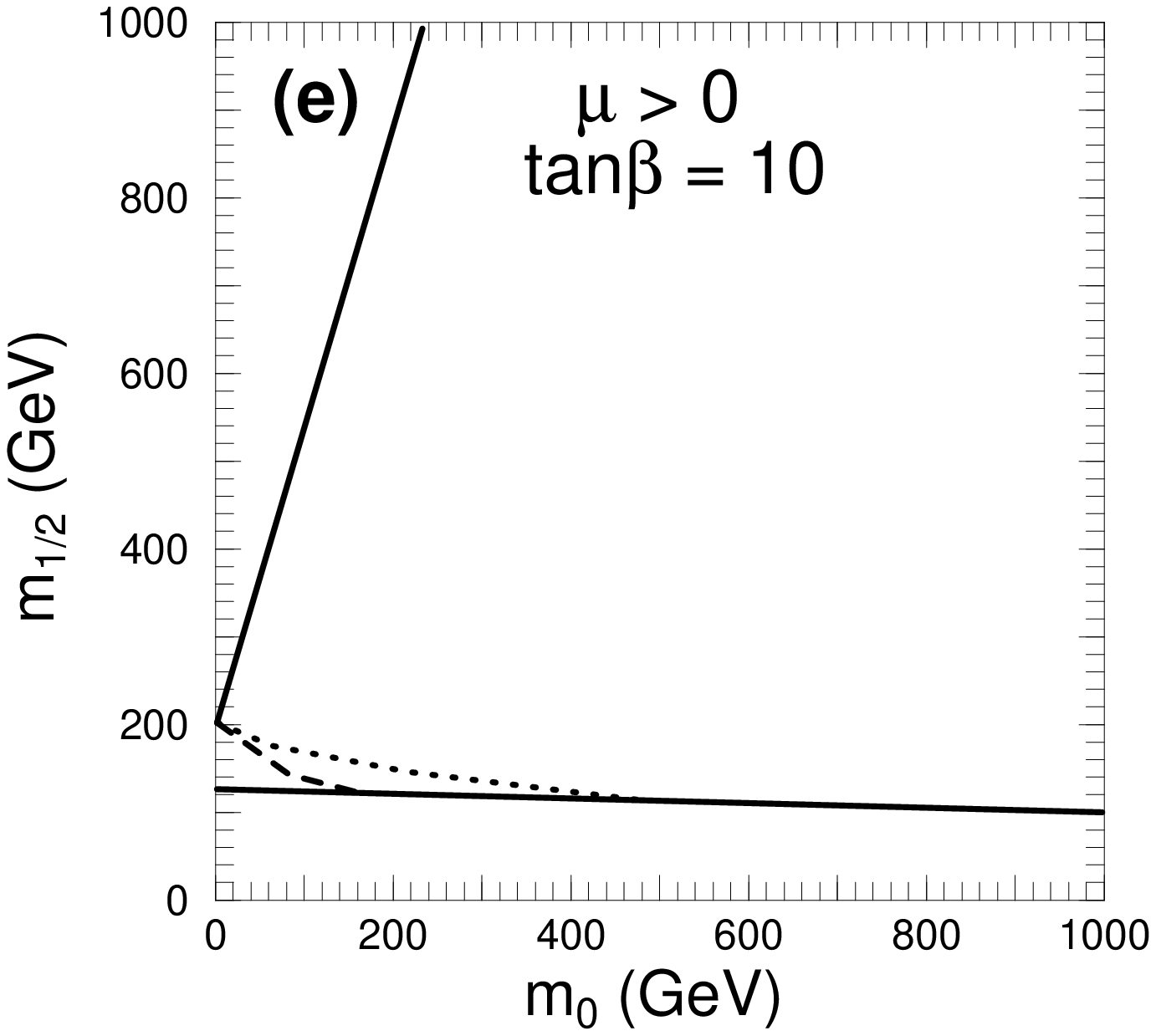,width=5.5cm,silent=0}
\leavevmode\hspace*{0.5cm}
\leavevmode\psfig{figure=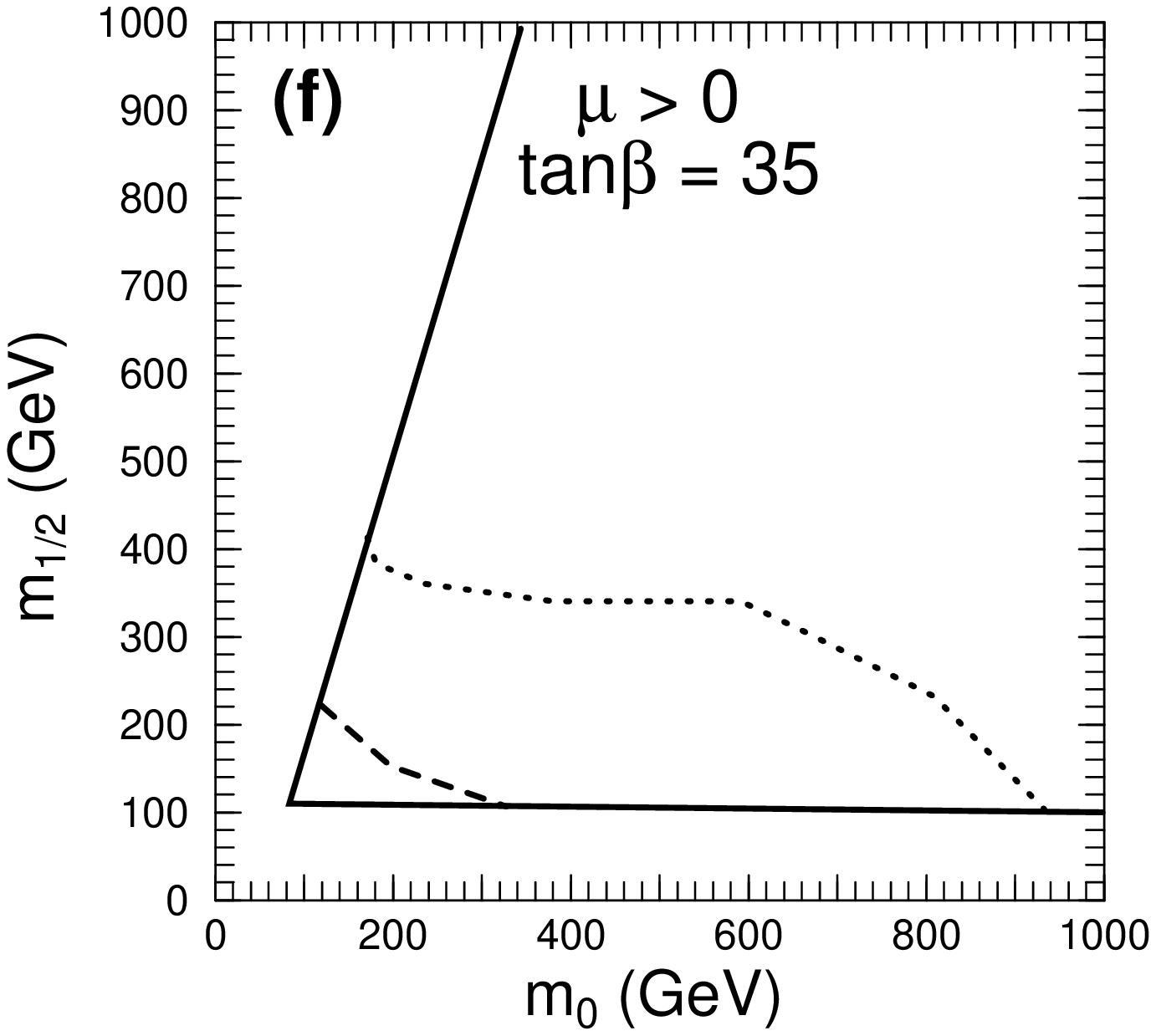,width=5.5cm,silent=0}
\end{center}
\caption{Favored regions in the mSUGRA $m_0$--$m_{1/2}$  plane lie in the 
region which is above and to the right of all drawn contours.  Further
explanation is provided in the text.}
\label{fig_m0-m1/2}
\end{figure}
Next we discuss Figures~\ref{fig_m0-m1/2}(a)--(f).  In each of these figures 
the values for $\tan\beta$ and ${\rm sign}(\mu)$ are held to the constant
values indicated.  We allow
$A_0$ to vary in the range $-$500~GeV $ < A_0 < $ 500~GeV, and we scan the 
$m_0$--$m_{{1/2}}$ plane between 0~GeV and 1~TeV.  For each point in the
five-dimensional parameter space of unification-scale input parameters we 
employ the mSUGRA RGE portion of ISAJET\cite{isajet7.40} to determine the 
RGE evolution to the electroweak scale.  We then verify whether that point is 
either excluded or allowed according to the following tests:
\begin{enumerate}
\item{Verify that the obtained particle spectrum is physical, that the correct 
vacuum for electroweak symmetry breaking is obtained and that the lightest 
superpartner particle is a neutralino, {\em i.e.} $\widetilde{\chi}^0_1$.  
This leads to a disallowed region in the upper left corner of each of the 
figures extending to the solid line with positive slope.\label{test-phys}}
\item{Verify that the chargino mass bound, $m_{\tilde{\chi}^\pm_1} > 
91$~GeV, is satisfied.  We find that region below the horizontal solid line is 
excluded.\label{test-mass}}
\item{Calculate $\Delta S^\prime$, $\Delta T^\prime$ and $\Delta m_W$ and 
check $\chi^2_{\rm tot}$.  Points which are disallowed at the 95\%~CL extend
the disallowed region in the $m_0$--$m_{{1/2}}$ plane from the solid
contour to the dashed contour.\label{test-stu}}
\item{Calculate the contribution to ${\rm Br}(B\rightarrow X_s\gamma)$.
Points which are disallowed at the 95\%~CL extend the disallowed region of the 
$m_0$--$m_{{1/2}}$ plane from the dashed contour up to the dotted 
contour.\label{test-bsg}}
\end{enumerate}
The portion of the $m_0$--$m_{{1/2}}$ plane which is above and to 
the right of all the contours is deemed the `favored' region for the mSUGRA 
model.  The portion of the $m_0$--$m_{{1/2}}$ plane which is excluded by 
Test~\ref{test-mass}, the chargino mass bound, is significant.  Once this has 
been taken into account, Test~\ref{test-stu} excludes a corner of the remaining
$m_0$--$m_{{1/2}}$ plane corresponding to small values of $m_0$ and 
$m_{{1/2}}$.  This region is fairly large in Figure~\ref{fig_m0-m1/2}(a)
while it is barely observable in Figure~\ref{fig_m0-m1/2}(d).  When 
${\rm sign}(\mu) < 0$, when $\tan\beta$ is large, and especially when both of these
conditions are true Test~\ref{test-bsg} excludes a significant region of the 
parameter space.  In Figure~\ref{fig_m0-m1/2}(c) all but a tiny portion of the
figure has been disallowed.  Our excluded regions from Test~\ref{test-bsg} 
are larger than those of Ref.~\cite{more-bsg} due to a different treatment of 
strong corrections.


In conclusion, the direct constraints which come from the nonobservation of 
the lightest chargino at LEP2 have important consequences.  First of all, the 
process dependent vertex and box corrections to four-fermion amplitudes become
negligibly small, and as a result the analysis of electroweak data has been 
simplified and has become more transparent.  After taking into account the 
chargino mass bound the $Z$-pole data, the low-energy neutral-current data and 
the measurement of the $W$-boson mass exclude only a small portion of the 
$m_0$--$m_{{1/2}}$ plane.  However, this is still significant because 
the excluded region is where $m_0$ and $m_{{1/2}}$ are small, precisely
the region of interest for collider studies, and especially relevant for the 
Tevatron.  We find that the excluded region is largest for smaller $\tan\beta$
with ${\rm sign}(\mu) < 0$.  For ${\rm sign}(\mu) < 0$ or $\tan\beta$ large,
a significant portion of the $m_0$--$m_{{1/2}}$ plane is excluded by the 
${\rm Br}(B\rightarrow X_s\gamma)$ measurement, and the constraint becomes 
very severe when both of these conditions are met.


\section*{Acknowledgments}
Conversations with Vernon Barger were especially useful for finalizing our 
report.  The work of Rob Szalapski is supported in part by U.S. Department of 
Energy under grant DE-FG02-91ER40685, and in part by the U.S. National Science 
Foundation under grants PHY-9600155 and INT9600243.  The research of Gi-Chol 
Cho is supported by Grant-in-Aid for Scientific Research from the Ministry of 
Education, Science and Culture of Japan.  Chung Kao has received support from 
the U.S. Department of Energy under grant DE-FG02-95ER40896 and from the 
University of Wisconsin Research Committee with funds granted by the Wisconsin
Alumni Research Foundation.


\end{document}